\documentclass{aa}  

\usepackage[varg]{txfonts}
\usepackage{graphicx}
\usepackage{orcidlink}
\newcommand{\co}{\mbox{\rm $^{12}$CO}}
\newcommand{\coa}{\mbox{\rm $^{13}$CO}}

\newcommand{\kms}{\mbox{\rm km s$^{-1}$}}
\newcommand{\htwo}{\mbox{\rm H$_2$}}

\newcommand{\xco}{\mbox{${\rm X}({\rm CO})$}}
\newcommand{\xcoup}{\mbox{${\rm X}({\rm CO})_{\rm upper}$}}

\newcommand{\cc}{\mbox{\rm cm$^{-3}$}}
\newcommand{\cmsq}{\mbox{\rm cm$^{-2}$}}
\newcommand{\msun}{\mbox{\rm M$_\odot$}}

\newcommand{\hi}{\mbox{H\,{\sc i}}}

\begin{document} 

\title{Cold molecular gas in the hot nuclear wind of the Milky Way}


\author{\orcidlink{0000-0002-3871-010X}M. Heyer\inst{1}
\and
\orcidlink{0000-0003-4019-0673}E. Di~Teodoro\inst{2}
\and 
\orcidlink{0000-0002-5635-3345}L. Loinard\inst{3,4,5}
\and 
\orcidlink{0000-0002-6050-2008}F.J. Lockman\inst{6}
\and 
\orcidlink{0000-0003-2730-957X}N.M. McClure-Griffiths\inst{7}
\and
\orcidlink{0000-0002-9279-4041}Q.D. Wang\inst{1}
}

\institute{Department of Astronomy, University of Massachusetts, Amherst, MA 01003, USA \\
\email{heyer@umass.edu} 
\and 
Dipartimento di Fisica e Astronomia, Universit\'a degli Studi di Firenze, via G. Sansone 1, I-50019 Sesto Fiorentino, Firenze, Italy 
\and
Instituto de Radioastronom\'{\i}a y Astrof\'{\i}sica, Universidad Nacional Auton\'oma de M\'exico, Apartado Postal 3-72, Morelia 58090, Michoac\'an, Mexico 
\and
Black Hole Initiative at Harvard University, 20 Garden Street, Cambridge, MA 02138, USA
\and
David Rockefeller Center for Latin American Studies, Harvard University, 1730 Cambridge Street, Cambridge, MA 02138, USA
\and 
National Radio Astronomy Observatory, Green Bank, WV 24944, USA
\and 
Research School of Astronomy and Astrophysics, Australian National University, ACT 2611, Australia
}

\date{Received XXX; accepted YYY}

 
\abstract{
Using the Large Millimeter Telescope and the SEQUOIA 
3~mm focal plane array, we 
have searched for molecular line emission from two atomic clouds associated with 
the Fermi Bubble of the Milky Way. 
Neither \co\ nor \coa\ J=1-0 emission is detected from the \hi\ cloud, MW-C20. 
\co\ J=1-0 emission is detected from MW-C21 that is distributed within 11 clumps 
with most of the CO luminosity coming from a single clump.  However, we 
find no \coa\ emission to a 3$\sigma$ brightness temperature limit of 0.3~K.
Using this limit and {\tt RADEX} non-local thermodynamic equilibrium (non-LTE) excitation models, 
we derive \htwo\ column density upper limits of (0.4-3)$\times$10$^{21}$ \cmsq\ for 
a set of physical conditions and a \htwo\ to \co\ abundance ratio of 10$^4$. 
Model CO-to-\htwo\ conversion factors are derived for each set of physical conditions.  
We find the maximum value is
1.6$\times$10$^{20}$ cm$^{-2}$/(K km s$^{-1})$.  Increasing [\htwo/\co] to 
10$^5$ to account for photodissociation and cosmic ray ionization increases the column 
density and \xco\ upper 
limits by a factor of 10.
Applying these \xco\ limits to the CO luminosities, 
the upper limit on the total molecular mass 
in MW-C21 is 132$\pm$2~\msun, corresponding to  $<$27\% of the neutral gas mass.
For the three clumps that are fully resolved, lower limits to the virial ratios are 
288$\pm$32, 68$\pm$28, and 157$\pm$39, which suggest that 
these structures are bound by external pressure to remain dynamically stable over the 
entrainment time of 2$\times$10$^6$ years or are being disrupted by shear and expansion over 
the clump crossing times of 3-8$\times$10$^5$ years.
The observations presented in this study add to the growing census of cold gas entrained within the 
Galactic Center wind. 
}


\maketitle

\nolinenumbers
\section{Introduction}
Cold gas embedded within hot, outflowing plasma is commonly observed in nearby and distant galaxies \citep{Veilleux:2020}.
The hot outflowing component is driven by clustered supernovae activity or an 
AGN-like process of gas accreting onto a 
supermassive black hole. The presence of cold gas in the hot, expanding plasma is primarily 
attributed to the entrainment of neutral atomic and 
molecular clouds that 
were originally 
located within the wind's expansion cone in the central region of the 
galaxy \citep{Aalto:2012, Walter:2017}.  An alternative explanation is 
in-situ formation of cold gas by the condensation of hot gas and the development of thermal instabilities
from which atoms and molecules can emerge \citep{Thompson:2016}.

Such multi-phase winds have an important role in the evolution of galaxies. The radiative and mechanical feedback 
can suppress star formation activity where the wind interacts with the cold, neutral gas of the galaxy. 
The extreme conditions of the entrained neutral gas are likely unsuitable for producing new stars.  
This suppression 
of star formation activity by feedback processes can account for the observed decrements of the 
galaxy luminosity function relative to 
theoretical expectations \citep{Silk:2012}.   The decrement at the low end of the luminosity function is attributed to supernova 
feedback while
AGN activity may be responsible for the decrement at the high end of the luminosity function.

The Fermi Bubble is a nearby example of an energetic galactic wind likely powered by an accretion event onto the 
massive Sgr~A* black hole \citep{Su:2010}.  
It is comprised
of
two symmetric plumes of hot electron gas that
extend 10 kpc above and below the disk.  \citet{Heywood:2019} found bipolar, non-thermal 
radio emission at 1.3~GHz that extends ~240~pc below the plane and 190~pc above the plane. 
X-ray
emitting lobes are located at the base of the plumes that indicate a
secondary, lower energy  outflow likely driven by the super star clusters in the central 200~pc of the Milky Way
\citep{Ponti:2019,Nakashima:2019}.  
In projection, neutral, atomic clouds are observed within the boundaries of the Fermi Bubble that indicate 
a multiphase
galactic wind
\citep{McClure-Griffiths:2013, DiTeodoro:2018}. 
Recent maps of \co\ J=2-1 emission 
have identified molecular gas clouds within the domains of two compact, atomic clouds  (MW-C1 and MW-C2) and
demonstrate that the multi-phase
outflow includes molecular gas
\citep{DiTeodoro:2020}.   A larger census of molecular line emission has been made in 17 additional atomic clouds (MW-C3 to MW-C19)
(Di~Teodoro, in preparation).

The multi-phase wind in the Milky Way affords an opportunity to investigate this hot-cold wind phenomenon with high spatial 
resolution.  Specifically, \hi\ 21~cm and CO spectroscopy can establish radial velocities of the entrained gas 
and a more detailed view of the cold gas kinematics. 
From the spectroscopic data, we can compile gas properties of the entrained clouds that can 
be compared to cold, neutral clouds located in the disk to assess the impact of the local environment 
on the gas 
conditions.

In this program, we extend the search for molecular gas in the Fermi Bubble by mapping 
\co\ and \coa\ J=1-0 emission from the atomic gas 
clouds G1.7+3.7-234 (hereafter, MW-C20) and  G358.7+3.7+179 (hereafter, MW-C21) that are identified in a blind survey of \hi\ 21cm line emission over 
the areas 0$^{\circ}$ $\le$ $|l|$ $\le$ 10$^{\circ}$ and
3$^{\circ}$ $\le$ $|b|$ $\le$ 10$^\circ$ using the Green Bank Telescope (GBT) \citep{DiTeodoro:2018}.
These 2 clouds are selected from this compilation on the basis of
having \hi\ column densities greater than 1.5$\times$10$^{19}$~\cmsq\ and compact, centrally condensed atomic gas distributions 
as measured by the 9\arcmin\ half-power beam width (HPBW) of the GBT. 
MW-C20 was first 
identified in this survey. 
MW-C21 was originally identified in the HIPASS survey that applied a new algorithm to 
search for high velocity 
clouds \citep{Putman:2002} and is also included in the compilation of atomic clouds 
by \citet{DiTeodoro:2018}.
Within this sample, MW-C21 stands out as having the largest \hi\ 21~cm full width half maximum (FWHM) line width of 41~\kms.  
In this study, we adopt distances of 7.62~kpc and 8.53~kpc for MW-C20 and MW-C21 
that are based on the simplified, kinematic wind model of \citet{DiTeodoro:2018}, which 
accounts for the displacement off the Galactic plane and distance from the Galactic Center. 

\section{Data}
Observations of \co\ and \coa\ J=1–0 emission were obtained simultaneously with the 50~meter 
Large Millimeter Telescope (LMT) Alfonso Serrano in June 2023 and March, April, and August 2024
using the 16 element focal plane array receiver SEQUOIA. The half-power beamwidths of the telescope at the line rest frequencies for 
\co\ (115.2712018 GHz) and \coa\ (110.2013541 GHz) are 12\arcsec\ and 13\arcsec\ respectively. The Wide-band Array Roach Enabled Spectrometer (WARES) 
was used to process the spectral information using the configuration with 400 MHz bandwidth and 97 kHz per spectral channel, 
which provides a velocity resolution of 0.25 \kms\ for \co\  and 0.27 \kms\ for \coa.
Data were calibrated by a chopper wheel that allowed switching between the sky and an ambient temperature load. The chopper wheel 
method introduces a fractional uncertainty of $\sim$10\% to the measured antenna temperatures \citep{Narayanan:2008}.
Routine pointing and focus measurements were made to ensure positional accuracy and optimal gain.

To cover the 15\arcmin$\times$15\arcmin\ area for each target cloud, we observed 4 submaps each 
covering 7.5\arcmin$\times$7.5\arcmin\ using 
On-the-Fly (OTF) mapping with scanning along the Galactic longitude  axis. 
All data were processed with the LMT spectral line pipeline package, {\tt lmtoy} \citep{Teuben:2024},  
which included calibration by the system temperature, zero-order baseline subtraction, and gridding of the 
spectra into a spectral line data cube,  weighted by both 
a spatial jinc function kernel and 1/$\sigma^2$, where $\sigma$ is the root mean square of antenna temperature 
values in velocity intervals where no signal is expected. 
The spatial pixel size of the data cube is 5.5\arcsec, which approximates the Nyquist sampling rates
for imaging \co\ J=1-0 and \coa\ J=1-0 emission.
The data are spectrally smoothed and resampled to 1~\kms\ resolution and channel width.
A main-beam efficiency of 0.6 is applied to convert the data from the T$_{\rm A}^*$ 
temperature scale to main-beam temperatures, T$_{\rm mb}$.  
The median rms sensitivities in main-beam temperature units at 1~\kms\ resolution are:
0.18~K (\co) and 0.11~K (\coa) for MW-C20 and 0.15 (\co) and 0.10 (\coa) for MW-C21. 
In sections where the 4 submaps overlap, the rms values are lower by $\sim$20\% relative to these median values.

\section{Results}
\begin{figure*}
\begin{center}
\includegraphics[width=0.85\textwidth]{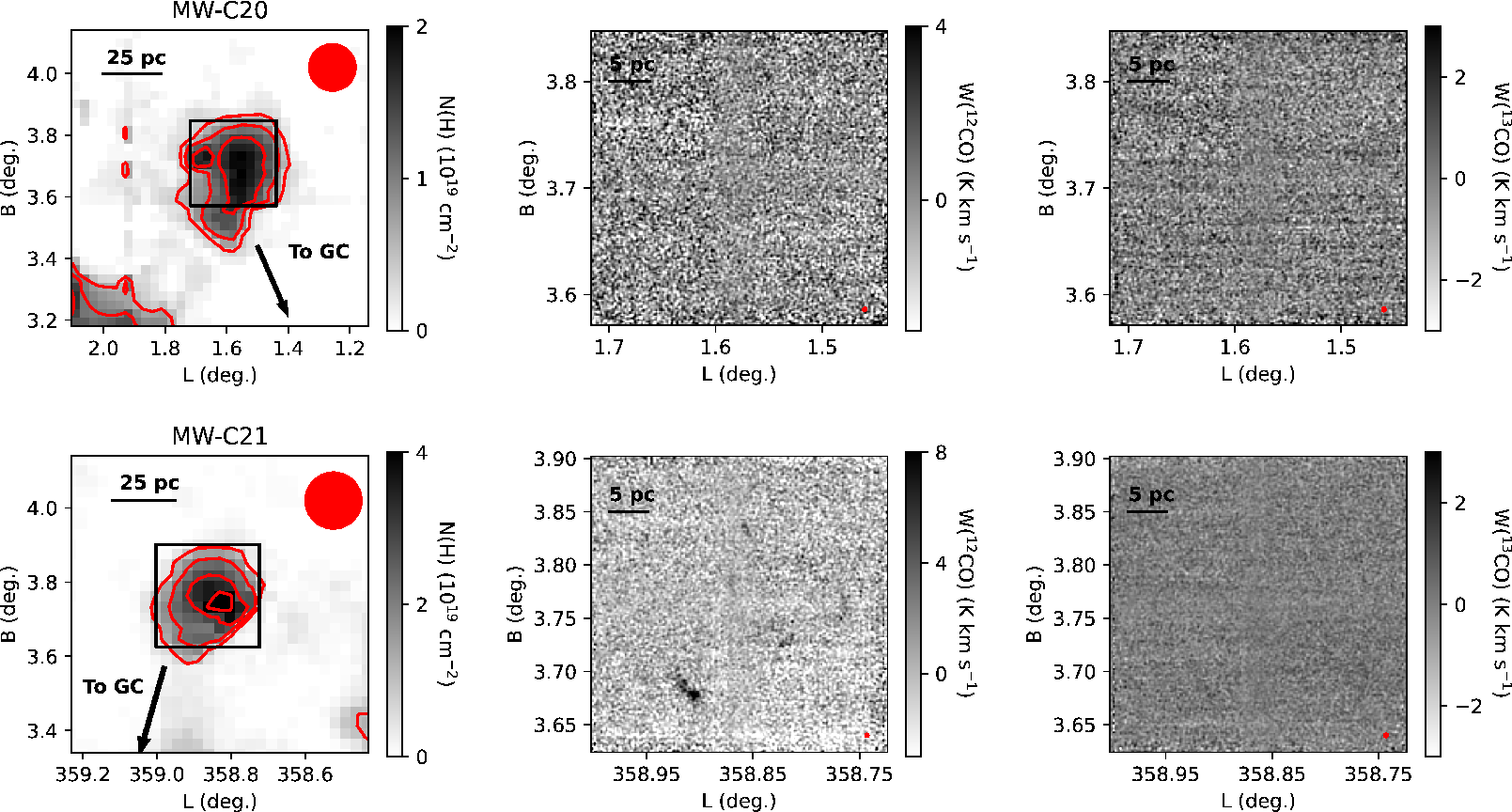}
\caption{Images of atomic hydrogen column density, \co\ and \coa\ integrated intensities for (top row) 
MW-C20 and (bottom row) MW-C21. 
The N(\hi) contour levels are (0.5, 1, 1.5)$\times$10$^{19}$~\cmsq\ and (1, 2, 3, 4)$\times$10$^{19}$ \cmsq, respectively. 
While \co\
emission is visible in several regions of MW-C21, the broad velocity range with respect to 
the narrow velocity components dilutes the signal with excess noise. 
No \coa\ J=1-0 emission is detected in either cloud.  The red circles show the GBT half power beam width 
of the HI observations.  The red circles in the lower right corner of the middle and right images denote 
the half power beam widths at \co\ and \coa. 
}
\label{figure1}
\end{center}
\end{figure*}

The compiled \co\ and \coa\ spectroscopic maps of MW-C20 and MW-C21 enable a search for molecular gas within the velocity range of the 
\hi\ 21~cm line emission.  For MW-C20, the \hi\ 21~cm line emission covers the velocity range of [-260,-200]~\kms\ 
with the strongest emission between [-250,-220]~\kms.  There is secondary \hi\ velocity component in the observed 
field that ranges from -160~\kms\ to -110~\kms\ that spatially overlaps with MW-C20.
The \hi\ emission from MW-C21  spans the velocity interval 130 to 225~\kms\ with the brightest emission  
between 150 and 190~\kms. 
Figure~\ref{figure1} shows the column density maps of atomic gas from \citet{DiTeodoro:2018},
\co\ and \coa\ J=1-0 maps of integrated intensity for both clouds over the velocity intervals [-250,-220]~\kms\  for 
MW-C20 and [150,190]~\kms\ for MW-C21. 
For MW-C20, \co\ and \coa\ velocity integrated emission is not detected to 
3$\sigma$ limits of 3.0 and 1.8 K~\kms\ respectively. Nor is either transition found in the 
secondary velocity component 
between -160~\kms\ and -110~\kms.
\co\ emission is 
detected in several regions within the MW-C21 atomic cloud.  
However, we find no detected \coa\ velocity integrated emission to a 3$\sigma$ limit of 1.9~K~\kms. 

An examination of the MW-C21 \co\ spectra shows there are 
several distinct velocity components within the cloud. 
The FWHM line widths are much narrower than the velocity interval 
over which the spectra are integrated.  This introduces excess noise to the spectra that can limit the detection of CO emission.  
To improve the signal to noise ratio (S2NR), the \co\ J=1-0 data cube is segmented into islands of detected emission using 
the {\tt pycprops} module, 
 which is a python implementation of the {\tt cprops} package \citep{Rosolowsky:2006}.
The root mean square (rms) is calculated for channels outside of the velocity interval [130, 190]~\kms\ for each pixel.
A S2NR data cube is generated by dividing the brightness temperature in each velocity channel by its rms value for all spatial pixels.
Islands (hereafter, clumps) are identified based on the 3-dimensional connectivity of 
neighboring spatial pixels and velocity channels   
with signal to noise ratios greater than 1.5. The choice of the S2NR threshold of 1.5 allows us to recover 
faint signal with respect to our sensitivity. To reduce the probability of false positive identifications, we also require that an 
island have at least 3 connected spectral channels that satisfies the S2NR threshold for each spectrum 
and be comprised of 4 or more spatial pixels.  
The probability of a random value sampled from a normal distribution being greater than 1.5$\sigma$ is 0.067.
The probability that 3 contiguous channels exceed 1.5$\sigma$ is (0.067)$^3$=0.0003.  The 
4 spatial
pixel minimum criterion further decreases the false positive probability. 
The clumps are not decomposed further into smaller objects. 

\begin{figure}
\begin{center}
\includegraphics[width=0.35\textwidth]{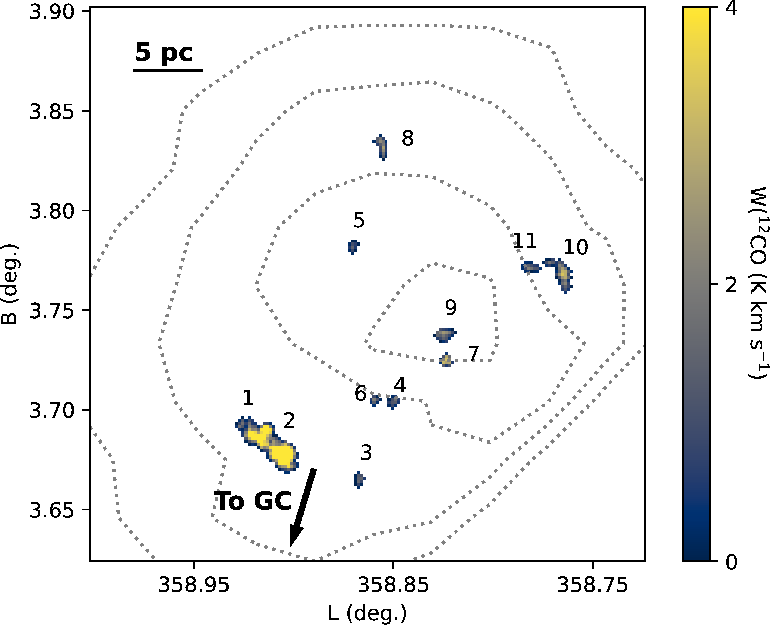}
\caption{Image of \co\ integrated intensities of clumps in MW-C21 
based on the 3 dimensional mask array that identifies islands of connected CO emission. 
The dotted contours show the distribution of atomic hydrogen column density with contour levels (1,2,3,4)$\times$10$^{19}$ \cmsq. 
}
\label{figure2}
\end{center}
\end{figure}

\begin{figure*}
\begin{center}
\includegraphics[width=0.9\textwidth]{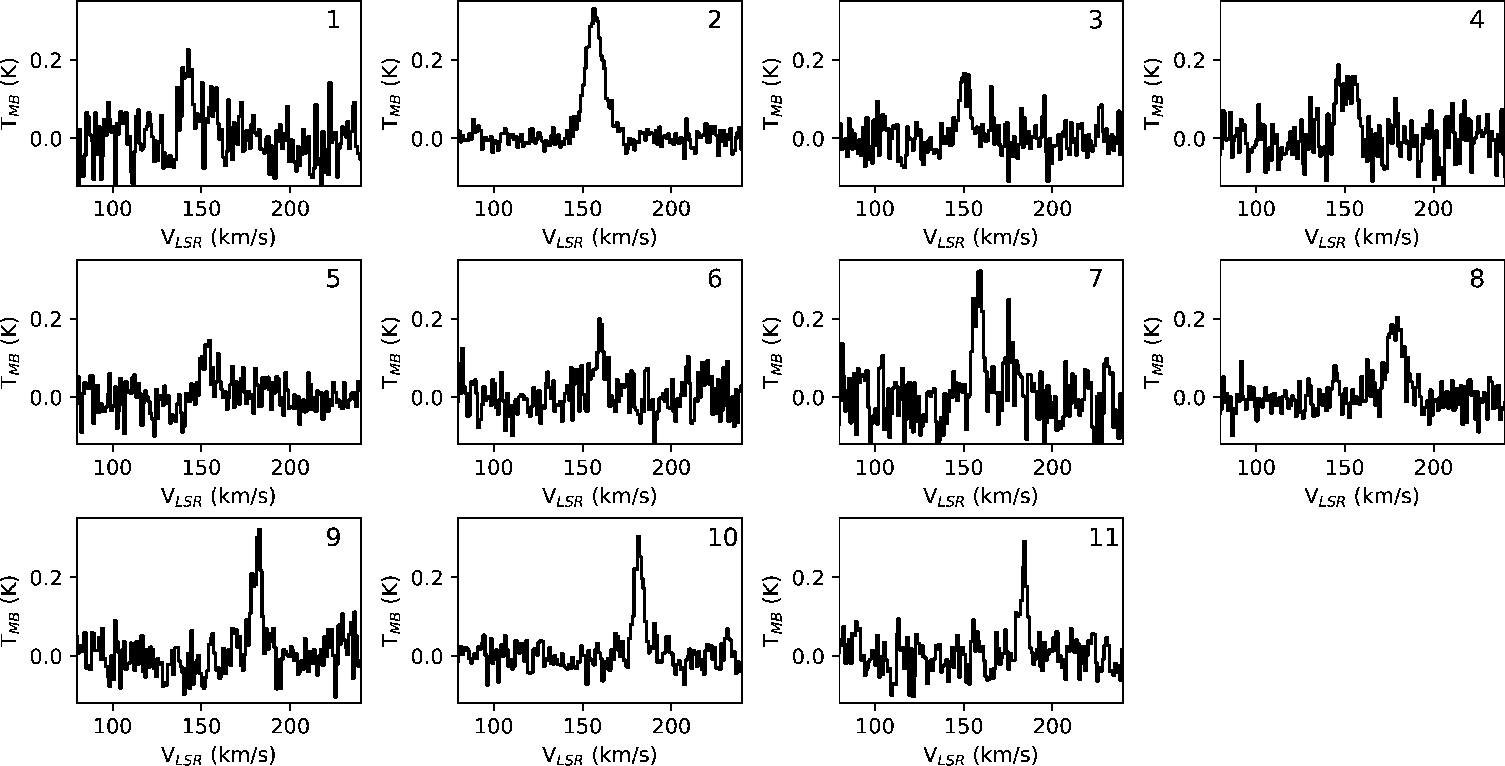}
\caption{Coadded average \co\ J=1-0 spectra for each molecular clump in MW-C21 based on the 3 dimensional mask array.
}
\label{figure3}
\end{center}
\end{figure*}

Applying this segmentation method to the MW-C21 \co\ data,  11 molecular clumps are identified in the observed 
field.
This decomposition provides a more precise, customized 
 velocity interval over which there is detected CO emission to improve the signal to noise of 
integrated intensity at the native 12\arcsec\ angular resolution for each labeled clump. The location and integrated intensity 
of the clumps are displayed in 
Figure~\ref{figure2}.  
The average spectra over the solid angle of each clump 
are shown in Figure~\ref{figure3}. 


We have further searched for \coa\ emission from MW-C21 by applying the \co-derived masks to the \coa\ data cube for each clump.  
Despite the reduction of noise given the customized velocity interval, \coa\ emission is still not rigorously detected in any 
of the clumps.  

The {\tt pycprops} module derives properties of the detected clumps.  
The calculation for each parameter is summarized by \citet{Rosolowsky:2006}.  Table~1 lists the number of 
position-position-velocity (ppv) voxels,  the 
positional centroid coordinates and centroid radial velocity, velocity dispersion (temperature weighted second moment), equivalent circular radius that 
matches the clump area, and CO luminosity.  
For the velocity dispersion, spatial moments ($\sigma_{r,maj}, \sigma_{r,min}$), and CO luminosity, {\tt pycprops} applies 
a correction to 
account for sensitivity bias.  \citet{Rosolowsky:2006} provide a detailed description for computing this correction.
The extrapolated spatial moments are deconvolved with the 
telescope beam dispersion width, $\sigma_{beam}$,
\begin{equation}
\sigma_r=\bigl[(\sigma_{r,maj}^2-\sigma_{beam}^2)^{1/2} (\sigma_{r,min}^2-\sigma_{beam}^2)^{1/2}\bigr]^{1/2}
\end{equation} 
\citep{Rosolowsky:2006}. 
Clumps with 
$\sigma_{r,min} < \sigma_{beam}$ are considered unresolved. 
Uncertainties for several of the clump properties are derived within the {\tt pycprops} package 
using the bootstrapping method in which a randomly selected 
subset of the full data sample are used to derive a given property \citep{efron:1979}.  This is repeated 128 times to generate a distribution of 
values from which confidence levels are determined. 
\begin{table*}
\label{table1}
\centering
\caption{Properties of molecular clumps in MW-C21}
\begin{tabular}{lcccccccccc}
\hline
\hline
ID & N Voxels & l & b & $V_{LSR}$ & $\sigma_v$ & $\sigma(\sigma_v)$ & R & $\sigma$(R) &  $L_{CO}$ & $\sigma(L_{CO})$  \\ 
 &  & (deg)  & (deg) & (km s$^{-1}$) & (km s$^{-1}$) & (km s$^{-1}$) & (pc)  & (pc) &  (K km s$^{-1}$ pc$^2$) & (K km s$^{-1}$ pc$^2$) \\ 
\hline
 1 &   40 & 358.926 &   3.692 &  142.7  &   1.3 &  0.3 & ...& ...&   0.5 &  0.1 \\ 
 2 & 1308 & 358.910 &   3.681 &  157.2  &   4.5  &  0.2 &  1.3 & 0.04 &  27.9 &  1.1 \\ 
 3 &   60 & 358.867 &   3.665 &  151.4  &   1.9 &  0.4 & ... & ... &   0.5 &  0.1 \\ 
 4 &   52 & 358.850 &   3.704 &  153.8  &   2.0 &  0.4 & ... & ... &   0.5 &  0.1 \\ 
 5 &   37 & 358.871 &   3.782 &  153.2  &   1.4 &  0.3 & ... & ... &   0.3 &  0.1 \\ 
 6 &   43 & 358.860 &   3.705 &  158.7  &   2.2 &  0.7 & ... & ... &   0.4 &  0.1 \\ 
 7 &   89 & 358.823 &   3.724 &  158.4  &   1.9 &  0.3 & ... & ... &   1.4 &  0.2 \\ 
 8 &  128 & 358.856 &   3.833 &  179.6  &   2.7 &  0.4 & ... & ... &   1.5 &  0.2 \\ 
 9 &   88 & 358.824 &   3.738 &  182.5  &   1.3  &  0.2 &  0.2 & 0.02 &   1.5 &  0.2 \\ 
10 &  231 & 358.765 &   3.768 &  182.4  &   1.7  &  0.2 &  0.6 & 0.05 &   3.4 &  0.3 \\ 
11 &   52 & 358.781 &   3.771 &  184.8  &   1.0 &  0.2 & ... & ... &   0.8 &  0.1 \\ 
\hline
\smallskip
\end{tabular}
\end{table*}

\begin{figure*}
\begin{center}
\includegraphics[width=0.7\textwidth]{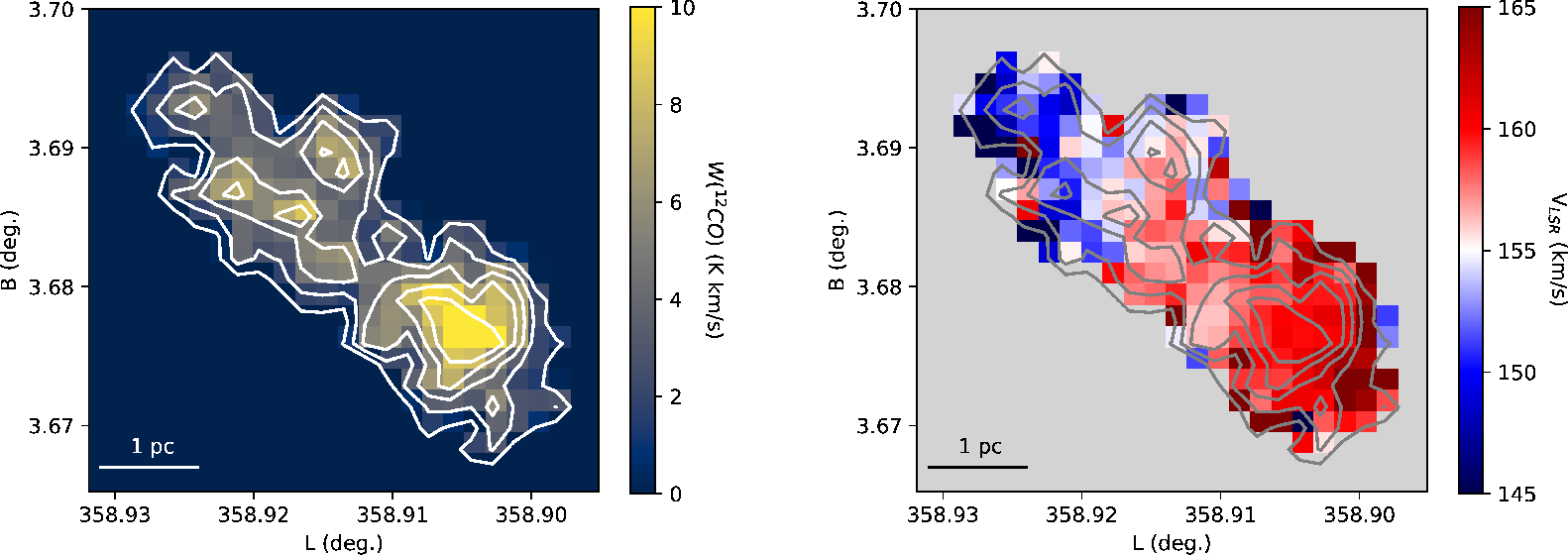}
\caption{Images of \co\ (left) integrated intensities and (right) centroid velocities that combine the masks for Clump~1 and Clump~2.
The contours show the distribution of \co\ integrated intensities with levels 1,3,5,7,9~K~\kms. 
}
\label{figure4}
\end{center}
\end{figure*}

The most conspicuous feature identified in the MW-C21 \co\ data is Clump~2, which exhibits
the largest size, velocity dispersion, and 
CO luminosity.  
Figure~\ref{figure4} shows the integrated intensity and 
velocity centroid images of Clump~2 and the adjacent Clump~1. 
While identified as a distinct object, Clump~1 is likely an extension of Clump~2 as 
its radial velocity smoothly 
extends the systematic 
shift from 
higher radial velocities in the lower right section of Clump~2 to lower velocities in the upper left section.
Measured end-to-end from Clump~2 to Clump~1, the radial velocity gradient is 1.3$\pm$0.4 \kms pc$^{-1}$. 
Added together, Clump~1 and Clump~2 
comprise 73\% of the detected CO luminosity within the field. 

The {\tt pycprops} algorithm was similarly applied to the MW-C20 \co\ J=1-0 data to search for 
any signal that could be diluted by excess noise in the integrated intensity image in Figure~\ref{figure1}. 
With a signal to noise threshold of 1.5, no \co\ emission could be found. 

\section{Discussion}
\subsection{Comparison of MW-C21 to MW-C1 and MW-C2}
The observations described in this study contribute one more example
to the 2 targets previously identified 
by \citet{DiTeodoro:2020} that contain 
 molecular gas within atomic clouds linked to the Fermi Bubble.
A comparison of MW-C21 to these  2 
clouds 
offers insight to the origin and nature of molecular gas in these environments.  

Morphologically, Clumps 1 and 2 form an extended wall of molecular gas within the MW-C21 atomic cloud 
that is oriented perpendicular to the direction of the Galactic Center  
and presumably, the flow direction of the hot wind. 
This orientation is similar to MW-C1 described by \citet{DiTeodoro:2020} and may be a signature to 
a cloud-wind shock front.  
Beyond this molecular gas wall, the remaining clumps 
are very compact and distributed downstream from Clumps 1 and 2.   This is similar to the 
\co\ distribution found in MW-C2. 

Kinematically, the \co\ velocity centroids of MW-C21 range from 142 to 185~\kms\ that are distributed in 2 distinct 
velocity intervals over a physical scale of 24~pc.  The higher velocity clumps with $V_{\rm LSR} >$ 175~\kms\ 
are compact and located 
downstream from Clump~2 along the direction to the Galactic Center and presumably, the wind flow direction.
This configuration suggests that the higher velocity clumps have been ablated from the largest molecular 
clump and accelerated by the galactic wind. 
MW-C2 also shows 2 velocity components that span 30~\kms\ over $\sim$20~pc scale and CO emission oriented 
along the direction to the Galactic Center.
For the individual velocity components in MW-C21, the velocity dispersions are between 0.9 and 4.5~\kms.  For comparison, 
the velocity dispersions are $\sim$1~\kms\ for MW-C1 and 2-5~\kms\ in MW-C2.  The velocity gradient identified in the 
composite velocity centroid image of Clumps 1 and 2  
indicates the presence of large scale velocity shear across the clump.  Velocity gradients are also found 
in MW-C1 and MW-C2 but 
are more 
localized.

\subsection{Clump masses}
Accurate values for the atomic and molecular masses are required to evaluate the dynamical state 
of the neutral gas.   Atomic gas column densities are 
derived from the integrated intensity of the \hi\ 21~cm emission over the cloud velocity interval, 
N(\hi)=1.823$\times$10$^{18}$ $\int {\rm T}_{\rm B,HI}(v)dv$ \cmsq\ that assumes 
optically thin emission.  The atomic mass of the cloud, M(\hi)=$\mu_{\rm H} {\rm m}_{\rm H}$D$^2$$\int$ N(HI) d$\Omega$,
where D is the distance to the cloud, $\mu_{\rm H}$ is the mean atomic weight=1.4 to account for the contributions from He, 
${\rm m}_{\rm H}$ is the mass of the hydrogen atom, 
  and the integration is over the solid angle of \hi\ emission. 
The masses of atomic 
gas within the 0.5$\times$10$^{19}$~\cmsq\ column density isophote are 257~\msun\ and 488~\msun\ for 
MW-C20 and MW-C21 respectively.  
The molecular hydrogen column density traced by \co\ J=1-0 emission is estimated using a CO-to-\htwo\ conversion factor, \xco,
\begin{equation}
{\rm N}({\rm H_2})={\rm X}_{\rm CO}{\rm W}(^{12}{\rm CO}) \;\;\; {\rm cm}^{-2}
\end{equation}
where ${\rm W}(^{12}{\rm CO})$ is the velocity integrated intensity of \co\ emission. 
\citet{DiTeodoro:2020} examined the range of possible values of \xco\ that could be applied to clouds within 
an extreme environment as the Fermi Bubble plumes using {\tt DESPOTIC} that considers the effects of the UV radiation field radiation field, $\chi$, and cosmic ray 
ionization rate $\zeta_{CR}$ on the CO abundance 
to predict line intensities \citep{Krumholz:2014}.
Given the observational constraints of ${\rm W}(^{12}{\rm CO})$ and cloud radius set by MW-C1 and MW-C2, 
they found a range of acceptable \xco\ values between 
2$\times$10$^{20}$~\cmsq~(K km s$^{-1}$)$^{-1}$, the average value in the Milky Way \citep{Bolatto:2013} 
and 40$\times$10$^{20}$ \cmsq (K km s$^{-1}$)$^{-1}$. 
This broad range of \xco\ values reflects the diversity 
of \co\ abundances that result from the applied model 
values of $\chi$ and $\zeta_{CR}$.

To corroborate the findings of \citet{DiTeodoro:2020}, 
we examine 
non-local thermodynamic (non-LTE) excitation models of \co\ and \coa\ brightness temperatures computed by 
{\tt RADEX} \citep{vanderTak:2007}. 
{\tt RADEX} generates brightness temperatures, excitation temperatures, and opacities 
for a given molecular line
as a
function of kinetic temperature, molecular column density, volume density, and line width that 
considers the population of molecular energy levels due to collisions of molecules
and radiative trapping.

While \coa\ J=1-0 emission is not detected in the 2 clouds, we can place upper limits on 
the molecular column density using the {\tt RADEX} models that assumes optically thin emission.   
A set of model \coa\ brightness 
temperatures are generated with the following set of physical conditions: 
kinetic temperature [10, 30, 100]~K, \htwo\ volume density [100, 316, 1000]~\cc, 
an \htwo\ to \co\ abundance ratio of 10$^4$,  
and 17 values of \co\ column density, N(\co),  
spaced logarithmically between 10$^{16}$~\cmsq\ and 
10$^{20}$~\cmsq.   
From these model values, the \coa\ column density required by {\tt RADEX} is 
N(\coa)=N(\co)[$^{13}$C/$^{12}$C]. 
We adopt the value of 
[$^{13}$C/$^{12}$C]=1/13.25 
as reported by \citet{Jacob:2020} for the Galactic Center that is the assumed origin of the molecular 
gas currently entrained in the wind.
 The \htwo\ column density is N(\htwo)=N(\co)[\htwo/\co].
An [\htwo/\co] value of $\approx$10$^4$ indicates that most 
carbon atoms in the gas phase are in the form of \co, while larger values imply neutral or 
singly ionized forms of atomic carbon 
due to photodissociation and ionization. 
The measured velocity dispersion of Clump~2 is 4.5~\kms, which corresponds to 
an 11~\kms\ width for a boxcar profile that is assumed by {\tt RADEX}. 

\begin{figure*}
\begin{center}
\includegraphics[width=0.8\textwidth]{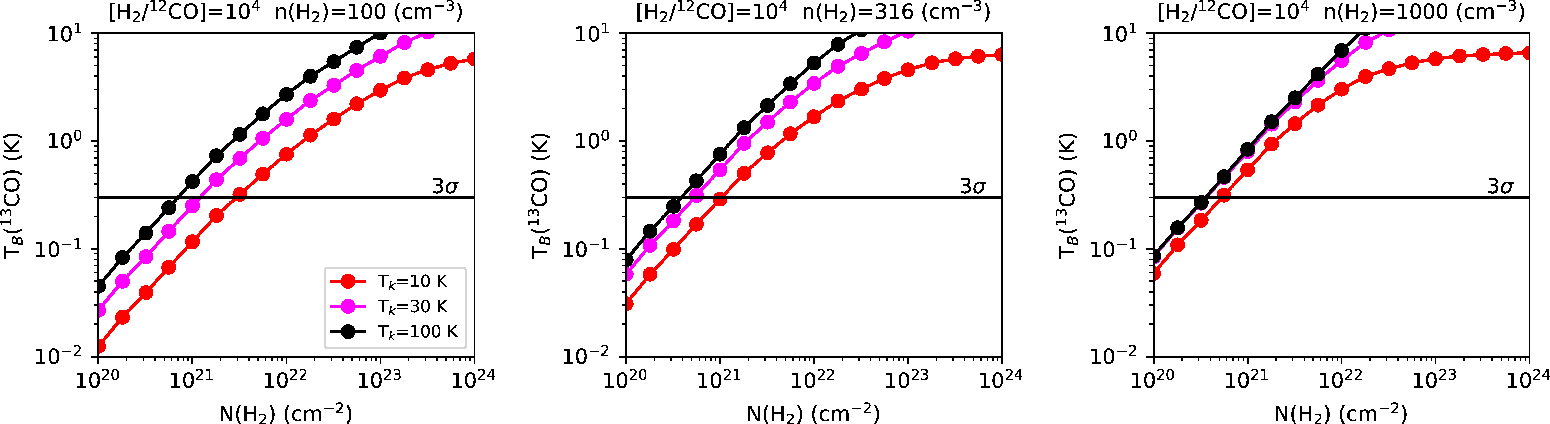}
\caption{Model \coa\ J=1-0 brightness temperatures as a function of \htwo\ column density for 
volume densities 100, 316, 1000 \cc, gas kinetic temperatures 10, 30, 100~K, 
[\htwo/\co]=10$^4$, 
and a line 
width of 11~\kms.  
We assume
a $^{12}$C to $^{13}$C abundance ratio of 13.25
\citep{Jacob:2020}. 
The horizontal black line marks the 3$\sigma$ threshold for the observed peak \coa\ brightness 
temperature from which upper limits on the \htwo\ column density for each model are estimated. 
}
\label{figure5}
\end{center}
\end{figure*}

The variation of \coa\ J=1-0 brightness temperature as a function of N(\htwo) is shown in 
Figure~\ref{figure5} for each set of kinetic temperatures and volume densities.
Also shown in Figure~\ref{figure5} is the 3$\sigma$=0.3~K brightness temperature 
upper limit of the \coa\ observations for MW-C21. 
For a given set of conditions, the intersection of this line with the model \coa\ brightness temperatures 
defines an upper 
limit for \htwo\ column density within the 13\arcsec\ beam of the observations.   These upper 
limits, N(\htwo)$_{\rm upper}$, range within (0.4-3)$\times$10$^{21}$ \cmsq\ depending on the 
selected kinetic temperature and 
volume density (see Table~2). 
\begin{table}
\label{table2}
\centering
\caption{Column density and \xco\ upper limits from RADEX models for [\htwo/\co]=10$^4$}
\begin{tabular}{ccccc}
\hline
\hline
n(\htwo) & T$_K$  & N(H$_2$)$_{\rm upper}$ & T$_{\rm B}$(\co) & \xcoup \\ 
(cm$^{-3}$) & (K) & (cm$^{-2}$)  & (K) & (cm$^{-2}$/(K km s$^{-1}$) \\ 
\hline
100 & 10 & 3.0e+21 &   1.7 &  1.6e+20\\
100 & 30 & 1.2e+21 &   2.0 &  5.4e+19\\
100 & 100 & 7.0e+20 &   2.4 &  2.7e+19\\
316 & 10 & 1.0e+21 &   1.9 &  5.0e+19\\
316 & 30 & 5.4e+20 &   2.6 &  1.9e+19\\
316 & 100 & 3.9e+20 &   3.1 &  1.1e+19\\
1000 & 10 & 5.3e+20 &   2.4 &  2.0e+19\\
1000 & 30 & 3.6e+20 &   3.3 &  1.0e+19\\
1000 & 100 & 3.6e+20 &   3.8 &  8.6e+18\\
\hline
\end{tabular}
\end{table}

Using the RADEX models, an upper limit to the conversion factor, \xcoup, can be derived that is consistent with the 
non-detection of \coa\ emission. For each set of parameters, the model \co\ brightness temperature, T$_{B}$(\co), is 
calculated at the 
\htwo\ column density upper limit. The model \co\ integrated intensity, ${\rm W}(^{12}{\rm CO})$, is 
the product of this brightness temperature 
and the line width of 11~\kms. The upper limit on \xco\ is \xcoup=${\rm N}({\rm H_2})_{\rm upper}/{\rm W}(^{12}{\rm CO})$.
The values for T$_{\rm B}$(\co) and \xcoup\ for each model are listed in Table~2.  

The maximum \xcoup\ value is 1.6$\times$10$^{20}$~\cmsq~(K~\kms)$^{-1}$
that is 
set by model with the lowest density (10 \cc) 
and kinetic temperature (10 K). 
This value only applies to the model parameter 
[\htwo/\co]=10$^4$.  Increasing [\htwo/\co] by a factor of 10 
to 10$^5$ would result in increasing 
the upper limits of column density and \xco\ by the same factor of 10.
The range of \xcoup\ values for [\htwo/\co]=[10$^4$, 10$^5$] 
are comparable to the range of \xco\ values found by \citet{DiTeodoro:2020} using {\tt DESPOTIC}
models and similarly, 
reflect our uncertainty of the UV radiation field and cosmic ray ionization rate in the environment 
of the Fermi Bubble.  
In the following analysis, we apply the low value of 
\xcoup=1.6$\times$10$^{20}$~\cmsq~(K~\kms)$^{-1}$
but acknowledge that there is a systematic positive error due to the uncertain 
value of [\htwo/\co]. We also emphasize 
that these derived conversion factors provide upper limits on the column density and mass of these clouds.

To derive clump mass upper limits, we multiply the extrapolated CO luminosity values by $\alpha_{CO}$, which is 
\xco\ in units of \msun/(K~\kms pc$^2$)$^{-1}$ and includes the mass contributions from Helium. 
For \xcoup=1.6$\times$10$^{20}$ and $\mu_H$=1.4, $\alpha_{CO}$=3.4 \msun (K~\kms pc$^2$)$^{-1}$. 
The total mass upper limit of all identified molecular clumps in MW-C21 
is 132$\pm$2 with 
Clump~2 contributing 95$\pm$2~\msun.  For the three clumps that are well resolved by the telescope beam (2, 9, 10), 
mass surface densities are calculated from the extrapolated CO luminosity
and radius, $\Sigma_{mol}=\alpha_{CO}L_{CO}/\pi R^2$ \msun/pc$^2$.  The mass surface densities 
are: (18$\pm$1)~\msun pc$^2$ (Clump 2), (42$\pm$11)~\msun pc$^2$ (Clump 9), and 
(9$\pm$1)~\msun pc$^2$ (Clump~10). 
While these values are similar to the surface densities of molecular clouds in the Solar neighborhood, 38 \msun/pc$^2$ \citep{Lewis:2022},
these 
are much lower 
than the surface densities found in clouds located in the Central Molecular Zone of the Galaxy ($\sim$1500  \msun/pc$^2$) from 
which the molecular features presumably became entrained in the wind flow
\citep{Oka:2001}.  The lower surface densities may indicate that the phase of neutral gas evolves 
over the entrainment time.

The molecular gas mass fraction defined as 
\begin{equation}
f{_{{\rm H}_2}}=\frac{{\rm M}({\rm H}_2)}{{\rm M}({\rm H}_2)+{\rm M}({\rm HI})}
\end{equation}
 has been used to evaluate the evolution of the entrained neutral gas 
phase \citep{Noon:2023}.
For MW-C21, the atomic gas 
mass within the area covered by the CO map 
is 353~\msun.  
The molecular gas fraction upper limit over the solid angle of the \co\ map is 0.27. 
\citet{DiTeodoro:2020} found $f{_{H_2}}$ values of 0.63 and 0.32 for MW-C1 and MW-C2 respectively. Applying a different mask,
\citet{Noon:2023} 
derived molecular gas fraction 0.53 and 0.44 for MW-C1 and MW-C2.  
Without higher angular resolution 21~cm~ data, we are unable to estimate the molecular gas fraction within each identified clump
as investigated by \citet{Noon:2023}. 
Both studies attribute the differences to 
differential exposure times to the contiguous 
hot wind.

\subsection{Virial ratio}
The dynamical state of a cloud  or clump is evaluated by the virial parameter, $\alpha_{vir}$, which is the ratio 
of the virial mass to the luminous mass as derived from spectral line or dust measurements.   
Clouds with values of $\alpha_{vir}$ $<$ 1 are self-gravitating objects.
Assuming the measured velocity dispersion represents random, turbulent
motions, clouds with virial ratios between 1 and 2 indicate 
equipartition between gravitational and kinetic energy densities.
If gas motions due to infall significantly contribute to the 
measured velocity dispersion, then $\alpha_{vir}$ values can vary between 1 and 2
\citep{Ballesteros:2011}.  
Measured virial ratios greater than 2 indicate external pressure-bounded clouds or regions that are 
unbound \citep{Bertoldi:1992}.   In practice, the application of the virial 
ratio to interstellar molecular clouds is challenging as the underlying 
assumptions and projection effects lead to large uncertainties in $\alpha_{vir}$ values \citep{Beaumont:2013}. 

From the {\tt pycprops} segmentation, three  clumps (2, 9, 10) are well resolved by the telescope beam, which 
allows
a computation of the virial mass and virial ratio.  The {\tt pycprops} calculation of the virial mass 
does not include surface terms.  The virial mass for clumps 2, 9, and 10
are (2.7$\pm0.3$)$\times$10$^4$~\msun, 340$\pm$137~\msun, and 1838$\pm$453~\msun\ respectively.  
The corresponding virial ratio lower limits are 288$\pm$32, 68$\pm$28, and 157$\pm$39.
Even with a value of \xco\ 10 times larger in the case of [\htwo/\co]=10$^5$, 
the virial ratios would still be greater than 6. 
These molecular clumps are 
not bound by self-gravity and are not in equipartition between kinetic 
and gravitational energy densities. 

If the clouds are not bound by external pressure, then the clumps  should expand and 
break up over several clump crossing times, where $\tau_{cross}=2R/\sigma_v$. 
The 
crossing times for the 3 resolved clumps are 6$\times$10$^5$ years (Clump~2), 3$\times$10$^5$ years (Clump~9), 
and 8$\times$10$^5$ years (Clump~10), which are 
$\sim$3-6  times shorter than the entrainment time of 1.9$\times$10$^6$ years 
estimated from the kinematic, biconical wind model of 
\citet{DiTeodoro:2018}.  
After several crossing time intervals, one would expect the gas to be broadly dispersed and 
more exposed to the 
interstellar radiation field leading to a reduced abundance of molecular gas. 
Yet, the molecular gas persist over these time scales in MW-C21, which argues for stable clumps 
that are pressure bounded at their surface.  

For a non-gravitationally bound system
to maintain quasi-static equilibrium, the energy density
must be balanced by an external pressure component at the surface. 
The energy density of a clump, $U_{cl}$,  
is $<\rho>3\sigma_v^2$, where $<\rho>$ is the mean mass volume density of the clump and the factor of 3
accounts for a 3 dimensional velocity dispersion.  Assuming 
a spherical clump, 
\begin{equation} 
U_{cl} = \frac{9{\rm M}\sigma_v^2}{4\pi{\rm R}^3} 
\end{equation}
where ${\rm M}$ is the clump mass. Using 
the values for Clump~2, the internal energy density is 4.2$\times$10$^{-10}$ dynes cm$^{-2}$.  
For Clumps 9 and 10, the energy densities are 5.5$\times$10$^{-10}$ and 
6.3$\times$10$^{-11}$ dynes cm$^{-2}$ respectively.  

The molecular clumps are embedded within a cloud of atomic gas with a mass of 428~\msun\ distributed 
within a radius of 
40~pc \citep{DiTeodoro:2018}. 
Assuming a spherical cloud,  
the mean volume density is 0.06~\cc.  
An upper limit on the gas kinetic temperature can be made assuming the 
measured HI~21~cm velocity dispersion from \citet{DiTeodoro:2018} is due to purely thermal motions, 
such that ${\rm T}_{\rm k}={\rm m}_{\rm H}\sigma_v^2/{\rm k}_{\rm B}$=3.5$\times$10$^4$~K, where ${\rm m}_H$ is the \hi\ 
atomic mass and ${\rm k_{\rm B}}$ is the Boltzmann constant. The upper limit on the 
 cloud thermal pressure, nT,  
is 2.2$\times$10$^3$~K~\cc= 4.4$\times$10$^{-13}$ dynes~\cmsq, which is insufficient to bound the internal 
pressure of the molecular clumps. Cosmic-ray pressure and magnetic field 
surface tension are possible bounding agents but the strength of each is not well defined in the 
local environment of MW-C21. 

\subsection{The role of photodissociation}
The molecular gas is exposed to the ambient FUV radiation field that can dissociate \htwo\ 
molecules. 
\citet{Noon:2023} find evidence for non-equilibrium hydrogen chemistry
in MW-C1, MW-C2, MW-C3 in which the molecular gas is photodissociated by the radiation field 
in the local environment due to insufficient column densities to self-shield \htwo\ molecules. 
They derive a dissociation time scale,
\begin{equation}
\tau_{diss}=7.0\Bigl(\frac{\Sigma_{mol}/\chi}{10\; \msun\; {\rm pc}^{-2}}\Bigr) \;\; {\rm Myr}
\end{equation}
where $\chi$ is the strength of the interstellar radiation field normalized to the Solar neighborhood value. 
For the 3 resolved clouds, $\tau_{diss}$ values are $13/\chi$ Myr (Clump 2), $29/\chi$ Myr (Clump 9), and 
$6.3/\chi$ Myr (Clump 10).  The value of $\chi$ in the Fermi Bubble is not well established but would have 
to be $>$ 10 to be comparable to the entrainment time for MW-C21 in order to have an impact on the 
amount of molecular gas.  MW-C21 is located $\sim$550~pc above the plane and well displaced 
from clusters of OB stars so one might expect low $\chi$ values ($<$ 3) in this environment.


\section{Conclusions}
We have searched for CO emission from 2 atomic gas clouds associated with the Fermi Bubble.  \co\ J=1-0 
emission is detected in the cloud MW-C21.  This emission is distributed in 11 discrete clumps. 
\co\ is not detected in the cloud MW-C20 and \coa\ J=1-0 emission 
is not detected in either cloud. Based on {\tt RADEX} models to predict \coa\ J=1-0 brightness temperatures
over a range of physical conditions, the absence of detected \coa\ emission constrains the \htwo\ column density 
to be less than 3$\times$10$^{21}$ \cmsq\ and  
the CO to \htwo\ conversion factor, \xco, to be $\le$ 1.6$\times$10$^{20}$ \cmsq (K~\kms)$^{-1}$, assuming [\htwo/\co]=10$^4$. 
Clump mass upper limits range from 1~\msun\ to 95~\msun\ with a total molecular 
mass $\le$ 132~\msun, which comprises $\le$27\%  of the neutral gas mass in the cloud. Mass surface density upper limits 
for the three  clumps that are well resolved by the 12\arcsec\ beam are 18, 42, and 9~\msun pc$^{-2}$.  The derived virial 
ratio lower limits for the three  resolved clumps are high (288, 68, 157), yet molecular gas is still present over 
the entrainment time scale, which suggests that these molecular features are bound 
by an external pressure component.

\begin{acknowledgements}
This work would not have been possible without the long-term financial support from the
Mexican Humanities, Science and Technology Funding Agency, CONAHCYT (Consejo
Nacional de Humanidades, Ciencias y Tecnolog\'{\i}as), and the US National Science
Foundation (NSF), as well as the Instituto Nacional de Astrof\'{\i}sica, \'Optica y
Electr\'onica (INAOE) and the University of Massachusetts, Amherst (UMass). The
operation of the LMT is currently funded by CONAHCYT grant \#297324 and NSF grant
\#2034318.
The data described in this paper include LMT observations conducted under the
scientific programs, 2023-S1-UM-16 and 2024-S1-MX-2.
The LMT welcomes acknowledgement of the scientific and technical support offered by
the LMT staff during the observations and generation of data products provided to the
authors. 

This work made use of Astropy:\footnote{http://www.astropy.org} a community-developed core Python package and an ecosystem of tools and resources for astronomy \citep{astropy:2013, astropy:2018, astropy:2022}.  
LL acknowledges the support of UNAM-DGAPA PAPIIT grants IN108324 and IN112820 and CONACYT-CF grant 263356.
E.D.T was supported by the European Research Council (ERC) under grant agreement \#10104075.
This research was partially funded by the Australian Government through an Australian Research 
Council Australian Laureate Fellowship (project number FL210100039 awarded to NM-G).
The NRAO is a facility of the National Science Foundation operated by Associated Universities, Inc.

\end{acknowledgements}

\bibliographystyle{aa}
\bibliography{fb.bib} 

\end{document}